\documentclass[twocolumn,english]{IEEEtran}
\usepackage[latin9]{inputenc}
\usepackage{xcolor}
\usepackage{pdfcolmk}
\usepackage{babel}
\usepackage{array}
\usepackage{float}
\usepackage{graphicx}
\PassOptionsToPackage{normalem}{ulem}
\usepackage{ulem}
\usepackage{subscript}
\usepackage[unicode=true,
 bookmarks=true,bookmarksnumbered=true,bookmarksopen=true,bookmarksopenlevel=1,
 breaklinks=false,pdfborder={0 0 0},pdfborderstyle={},backref=false,colorlinks=false]
 {hyperref}
\hypersetup{pdftitle={Your Title},
 pdfauthor={Your Name},
 pdfpagelayout=OneColumn,pdfnewwindow=true,pdfstartview=XYZ,plainpages=false}

\makeatletter

\providecommand{\tabularnewline}{\\}
\floatstyle{ruled}
\newfloat{algorithm}{tbp}{loa}
\providecommand{\algorithmname}{Algorithm}
\floatname{algorithm}{\protect\algorithmname}
\providecolor{lyxadded}{rgb}{0,0,1}
\providecolor{lyxdeleted}{rgb}{1,0,0}

\@ifundefined{definecolor}
 {\usepackage{color}}{}
\usepackage{babel}
\usepackage{array}

\providecommand{\tabularnewline}{\\}

 \let\oldforeign@language\foreign@language
 \DeclareRobustCommand{\foreign@language}[1]{%
   \lowercase{\oldforeign@language{#1}}}

\ifCLASSOPTIONcompsoc
\else
\fi
\PassOptionsToPackage{warn}{textcomp}

\makeatother

\begin{document}

\title{Neutron induced strike: On the likelihood of multiple bit-flips in
logic circuits}

\author{Nanditha P. Rao and Madhav P. Desai\\
Indian Institute of Technology Bombay\\
Email: \{nanditha@ee, madhav@ee\}.iitb.ac.in}
\maketitle
\begin{abstract}
High energy particles from cosmic rays or packaging materials can
generate a glitch or a current transient (single event transient or
SET) in a logic circuit. This SET can eventually get captured in a
register resulting in a flip of the register content, which is known
as soft error or single-event upset (SEU). A soft error is typically
modeled as a probabilistic single bit-flip model. In developing such
abstract fault models, an important issue to consider is the likelihood
of multiple bit errors caused by particle strikes. The fact that an
SET causes multiple flips is noted in the literature. We perform a
characterization study of the impact of an SET on a logic circuit
to quantify the extent to which an SET can cause multiple bit flips.
We use post-layout circuit simulations and Monte Carlo sampling scheme
to get accurate bit-flip statistics. We perform our simulations on
ISCAS'85, ISCAS'89 and ITC'99 benchmarks in 180nm and 65nm technologies.
We find that a substantial fraction of SEU outcomes had multiple register
flips. We futher analyse the individual contributions of the strike
on a register and the strike on a logic gate, to multiple flips. We
find that, amongst the erroneous outcomes, the probability of multiple
bit-flips for \textquoteleft gate-strike\textquoteright{} cases was
substantial and went up to 50\%, where as those for \textquoteleft register-strike\textquoteright{}
cases was just about 2\%. This implies that, in principle, we can
eliminate the flips due to register strikes using hardened flip-flop
designs. However, in such designs, out of the remaining flips which
will be due to gate strikes, a large fraction is likely to be multiple
flips.
\end{abstract}


\section{\label{sec:Introduction}Introduction}

\IEEEPARstart{ S}{oft} errors are known to have a significant
impact on circuit reliability. High energy particles from either cosmic
rays or packaging materials are the major contributors towards soft
errors. When such particles strike a semiconductor substrate, they
generate charge which in turn results in a glitch or a transient current
in a circuit. Such a glitch is called the single event transient (SET).
The SET can occur on a register or on a logic gate and can propagate
to eventually get captured in a register, altering the stored bit
($0\rightarrow1\,or\,1\rightarrow0$). Such an error is known as a
single event upset (SEU) or soft error. The cosmic rays interact with
earth's atmosphere creating secondary particles, mainly neutrons,
protons, muons and pions, as they penetrate down to the sea level.
The energy or flux of these particles increases at a rate of 5x per
5000 feet, reaching a maximum intensity at about 12-15 km from sea
level \cite{IBM_terrestrial,alt_lat_variations}. Thus, the errors
caused due to these particles were of specific interest in satellite,
aircraft and space applications. For example, in-flight measurements
reveal an upset rate of approximately 5x10\textsuperscript{-3} upsets
per hour per memory device \cite{in-flight}. 

Several studies were then conducted by IBM and were also summarized
by Boeing Defense and Space group, which reveal that a significant
amount of neutron flux is found even at sea level \cite{IBM_terrestrial,gnd_level}.
The average low energy neutron flux (around 10MeV) at sea level is
reported to be nearly 100000 neutrons/sq.cm per year \cite{IBM_terrestrial}.
They report that thousands of upsets happen every year at the ground
level, which are mainly recorded by computer systems having error
detection and correction logs. For instance, the average error rate
in memories observed in several computers were reported to be nearly
1.5e-12 upsets/bit-hr and most of the upsets are reported to be due
to atmospheric neutrons \cite{gnd_level}. Thus, SEUs caused due to
neutrons gained significant importance. Most early studies focused
on SEUs only in memories. SEUs in logic circuits were not considered
because such circuits exhibit inherent masking phenomena, which prevented
the SETs from getting captured in a register/flip-flop. However, as
technology scales, the impact of masking phenomena tends to reduce
\cite{combn1_glitch_ht,MBU_combn,combn_seq,SER_trends3_combn} and
it is important to study SEUs in logic circuits.\textcolor{black}{{} }

At the architectural level, soft errors are commonly modeled by a
probabilistic single bit-flip fault model\textcolor{black}{{} }\cite{single_bit1,single_bit_flip_element,single_flip_arch,ReStore_single-bitflip}.
In developing such abstract fault models, an important issue to consider
is the likelihood of multiple bit errors caused by particle strikes.
This likelihood has been studied to a great extent in memories, but
has not been understood to the same extent in logic circuits. This
model has been challenged in \cite{correlations1,multiple_errors_archlevel,correlations3_mult_upsets,multibit_gatemodel},
which report that multiple bit flips do occur in logic circuits. However,
the state-of-art fault model continues to be a single bit-flip, single
cycle model for soft error in logic circuits. Reliability estimates
(such as mean time to failure/MTTF) and reliability enhancement techniques
(such as error correction codes etc) are also based on the assumption
that a single bit flip occurs due to a particle strike. However, in
reality, if multiple errors occur, these MTTF estimates are likely
to be optimistic and the error correction methods are likely to be
insufficient. Most of the existing techniques that estimate soft error
rate (SER) use approximate modeling techniques to arrive at these
conclusions which reduces the accuracy of their results \cite{seq1_algo,observ_algo,BDD_SER_Miskov-Zivanov,mars_SER_algo_hardng,SEAT_LA_SER_estimation}.
Therefore it is essential that our estimation technique be as accurate
as possible, in order to increase the confidence in the conclusions
on the multiple bit flip probability in logic circuits. \textcolor{black}{So,
we perform }a detailed characterization of the impact of an SET using
post-layout circuit (SPICE) simulations\textcolor{black}{{} and Monte
Carlo sampling scheme in order to get accurate bit-flip statistics.
Our goal is to quantify the extent to which the single bit-flip fault
model} is accurate. 

We performed a basic characterization to this extent in an earlier
work \cite{nanditha_paper1}. We evaluate the bit-flip statistics
by comparing the SET- induced circuit simulation with a fault-free
register-transfer-level (RTL) reference simulation. In our simulations,
we assume that an SET affects a single transistor \cite{double_exp_drain_cur1,drain_cur2}.
We run our experiments on the \textit{ISCAS'85, ISACAS'89} and \textit{ITC}'99
benchmark circuits in 180nm and 65nm technologies. We found that the
impact of an SET in a circuit can be understood as a two-cycle phenomenon,
that is, the SEU outcomes need to be observed across two clock cycles
(the cycle which had the SET injected and the following clock cycle)
in order to accurately capture the phenomenon. This leads us to the
fact that there are several possible SEU outcomes. We estimated the
relative probabilities of all SEU outcomes and the conditional probability
of multiple flips given that there is at least one error. In other
words, given that an SET propagates and causes an error in a flip-flop
in either of the two clock cycles, what is the probability that it
can flip multiple flip-flops? We analyse these results further and
evaluate this probability separately for strikes on logic gates and
strikes on registers to understand their individual contributions. 

We find that, overall, up to 8\% of the erroneous outcomes result
in multiple bit-flips. Although this probability is low, it can a
have significant impact on error-detection or correction schemes.
It means that a single bit error correction scheme can go wrong in
as much as 8\% of the cases, and these errors will go down as silent
undetected errors. The probability of multiple errors also increases
as technology is scaled (based on 180nm and 65nm data).\textcolor{black}{{}
}A key observation is that, amongst the erroneous outcomes, the probability
of multiple bit-flips for \textquoteleft gate-strike\textquoteright{}
cases was substantial and went up to 50\%, that is, these errors are
caused due to the propagation of the SET from the logic gate to the
flip-flop. On the other hand, out of the erroneous outcomes, the likelihood
of multiple flips for \textquoteleft register-strike\textquoteright{}
cases was just about 2\%. This implies that, if we were to do hardened
flip-flop designs to eliminate the flips due to register strikes,
then in such designs, out of the remaining flips which will be due
to `gate strikes', a large fraction will be multiple flips. So, although
the traditional circuit designs with hardened flip-flops will solve
one problem, they will uncover a different problem. 

Thus, our study reveals that multiple flips are quite likely and are
likely to increase with technology scaling. Reliability estimation/enhancement
approaches based on the single bit-flip model are likely to be optimistic.
Gate-strikes are the key contributors to multiple flips. Robust flip-flop
designs may not help; we may need to look at methods such as modifying
the path delays or designs such as delay-capture flip-flops to reduce
the likelihood of multiple errors due to gate-strikes. Enough precautions
need to be taken at the layout/circuit/system level so that the single
bit-flip model can be used with a higher degree of confidence.

We organize the rest of the paper as follows. In Section \ref{sec:TwoCyclePhenomenon},
we provide a brief introduction to the two-cycle phenomenon and describe
the possible SEU outcomes in a logic circuit. We describe the experimental
setup in Section \ref{sec:Methodology}. We present our simulation
results on \textit{ISCAS'85, ISCAS'89} and\textit{ ITC}'99 benchmarks
in Section \ref{sec:Results}. In Section \ref{sec:Conclusions},
we summarize our paper. 

\section{\label{sec:TwoCyclePhenomenon}Two-cycle phenomenon and the possible
SEU outcomes}

\begin{figure}
\includegraphics[scale=0.35]{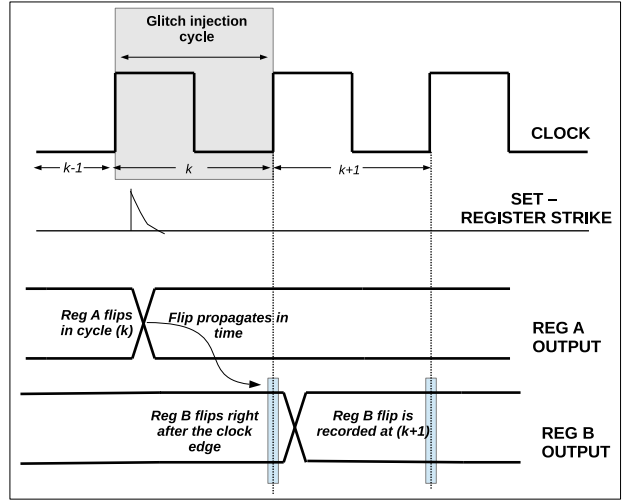}

\caption{\label{fig:FF-type}A flip occurs in both clock cycles due to the
SET occurring early in the clock cycle}
\end{figure}

\begin{figure}
\includegraphics[scale=0.35]{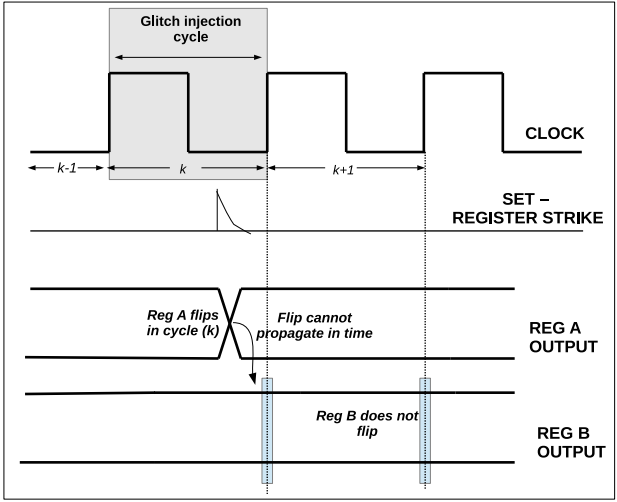}

\caption{\label{fig:FN-type}A flip does not propagate in time to the next
cycle due to the SET occurring late in the clock cycle}
\end{figure}

In a logic circuit, a high-energy particle can strike a logic gate
or a register, resulting in an SET. We call this the `gate-strike'
and `register-strike' respectively. \textcolor{black}{We model the
SET as a current injection at the drain of a transistor in a particular
clock cycle `k'. This SET can eventually propagate and get captured
in a register or a flip-flop in the same clock cycle `k' or in the
subsequent clock cycle `k+1', depending on the time instant at which
the SET occurs in clock cycle `k'. }These two possibilities can be
explained as follows. If the SET flips a register content early in
the clock cycle `k', it will have enough time to\textcolor{black}{{}
propagate and flip some register in clock cycle`k+1', as illustrated
in Figure \ref{fig:FF-type}. However, if the SET flips a register
later in the clock cycle `k', the error will not have enough time
to propagate and flip some other register in the subsequent clock
cycle `k+1', as illustrated in Figure \ref{fig:FN-type}. }Thus, the
impact of an SET, in reality, needs to be viewed across two clock
cycles to accurately capture the phenomenon. We call this the `two-cycle
phenomenon'. This time dependence of the strike is currently missing
in the single bit flip, single cycle fault model.\textcolor{black}{{} }

\begin{figure}[h]
\textcolor{black}{\includegraphics[scale=0.35]{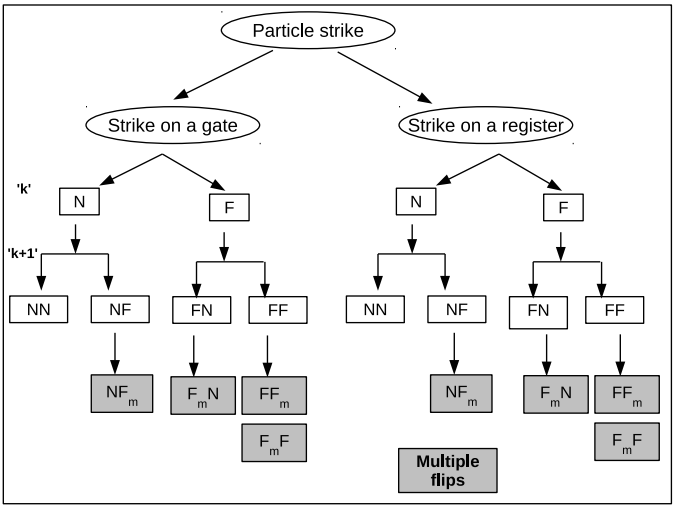}}

\caption{\label{fig:taxonomy}A systematic notation for the possible error
(SEU) outcomes caused due to an SET in a logic circuit}
\end{figure}

\textcolor{black}{We now classify the flips and come up with a systematic
notation for the possible SEU outcomes across two clock cycles. This
is shown in Figure \ref{fig:taxonomy}. In the figure, `N' stands
for no-flip, `F' stands for flip and }`F\textsubscript{m}\textcolor{black}{'
stands for multiple flips. So, the illustrations in Figure \ref{fig:FF-type}
and \ref{fig:FN-type} can be classified as `FF' and `FN' type of
outcomes respectively as per this terminology. }Thus, the occurrence
of second flip depends on the time instant at which the SET occurred
in the first clock cycle. Flips in the subsequent clock cycles can
be understood completely based on logical propagation of the flip
in the second cycle which occurs on the clock edge. 

\textcolor{black}{Since we are interested in evaluating the accuracy
of the single bit-flip model, we focus on the likelihood of multiple
register flips. The ones caused directly as a result of the SET are
denoted by `}NF\textsubscript{m}'\textcolor{black}{, }`F\textsubscript{m}F'\textcolor{black}{{}
and }`F\textsubscript{m}N'\textcolor{black}{. The multiple flip caused
due to the propagation of the previous flip is denoted by `}FF\textsubscript{m}',
which will be anyway captured by modeling the first flip using the
traditional single bit-flip model. So, the multiple flips in this
case are not a direct result of the SET and we do not consider this
in our experiments.

\section{\label{sec:Methodology}Simulation setup}

We perform our simulations on the \textit{ISCAS'85, ISCAS'89} and
\textit{ITC'99} benchmark circuits in 180nm and 65nm technologies.
We model the SET as a current injection at the drain of a single transistor
\cite{double_exp_drain_cur1,drain_cur2}. A fixed glitch is used in
our experiments and we assume that the probability that an SET affects
a drain is proportional to its area. We use the scaling trends presented
in \cite{dodd_prod_prop} to arrive at the glitch height and width
for the technology we simulate. 

Each circuit that we perform our experiment on, is described as a
Verilog or VHDL netlist (RTL). The circuit is implemented to layout
using synthesis and placement-route (PNR) tools. The post-layout Verilog
and circuit netlist are then extracted. We simulate the post-layout
Verilog netlist with a representative test-bench using the \textit{ModelSim}
\cite{modelsim} simulator and store the input/output values of all
the flip-flops (registers). These are called as the reference values.
From the post-layout circuit netlist, we generate `sample circuit
simulation decks' by picking a random clock cycle (k) for simulation,
selecting a random drain (d) to inject the SET and selecting a random
time instant (t) in the clock cycle to inject the SET. 

The sample circuit is simulated for three clock cycles as shown in
Figure \ref{fig:FN-type} and the register outputs are recorded at
two clock instants: `k' and `k+1'. We use \textcolor{black}{the Ngspice
or Hspice circuit simulators \cite{ngspice} to perform our simulations.}
In the (k-1)\textsuperscript{th} clock cycle, the input registers
of the circuit netlist are initialized with the \textcolor{black}{corresponding
input} values obtained from the RTL reference values \textcolor{black}{for
that clock cycle}. In the k\textsuperscript{th} clock cycle, we inject
the SET at a random point in time (t) in the clock cycle. Inputs to
the circuit change at the falling edge of the clock cycle k, so that
there is sufficient setup time. We note the register outputs at the
rising edges of clock cycles $k$ and $k+1$. The register values
from the SET injected circuit simulation are compared with the corresponding
reference values from the fault-free RTL simulation. Differences between
the sampled values in the circuit simulation and the reference values
are recorded as bit-flips in the respective clock cycles. 

We generate several sample simulation decks with different \textit{\textcolor{black}{(d,k,t)}}\textcolor{black}{{}
values. }The average number of sample circuit simulations run for
each circuit was about 4000. We ran these simulations in parallel
using GNU Parallel \cite{Tange2011a} on a high performance computing
cluster in Centre for Development of Advanced Computing (CDAC) Pune,
India, which utilizes multiple cores. The time taken to run these
many simulations for each circuit was on an average about 1 to 2 hours.
The key advantage of running the post-layout circuit simulations is
that they capture all the masking phenomena accurately. With the availability
of the high performance cluster with multi-core facility, these simulations
can be run in a reasonable amount of time. 

We extract the bit-flip information from each simulation and then
classify these flips into one of the SEU outcomes described in Figure
\ref{fig:taxonomy}. We continue to generate the circuit samples until
the standard error ($\frac{Standard\:deviation}{\sqrt{N}}$) is reduced
to less than 10\% of the value of the estimate. The probabilities
obtained from this Monte Carlo sampling experiment fall within the
95\% confidence interval. The experimental setup is briefly described
in Algorithm \ref{alg:Brief-flow-of}. This entire process is automated
using a set of python and perl scripts.

\begin{algorithm}
\textbf{Input: }Verilog/VHDL description of circuit
\begin{itemize}
\item Perform Synthesis of the Verilog/VHDL description
\item Perform placement and route of the resulting Synthesis output
\item Extract the post-layout Verilog and SPICE netlists
\item Simulate the post-layout Verilog netlist using ModelSim for 10000
clock cycles
\item Store the input and output values of all flip-flops at every clock
cycle
\item \textbf{for }(Number of simulations)

\{
\item Create a sample simulation netlist from the post-layout SPICE netlist:

$\;\;\;$\{
\begin{itemize}
\item Pick a random drain \textit{d} of a transistor at which the SET will
be injected
\item Pick a random clock cycle \textit{k} from the reference trace for
simulation
\item Pick the inputs to the sample netlist from the chosen clock cycle
\textit{k}
\item Inject the SET at a random time instant \textit{t} in the selected
clock cycle
\end{itemize}
$\;\;\;$\}
\item Simulate the sample SPICE netlist with parameters $d,k,t$ for 3 clock
cycles
\item Store the output values of all flip-flops at the 2nd and 3rd clock
edges
\item Compare the outputs from SPICE netlist with that of the Verilog netlist
\item Note any discrepancy in the output as a bit-flip
\item Evaluate the conditional probability of multiple flips given at least
one error occured
\item Simulations are run to reach a 95\% confidence interval

\}
\end{itemize}
\textbf{Output:} Calculate the statistics of all SEU outcomes

\caption{\label{alg:Brief-flow-of}Experimental setup to perform the circuit
level characterization of the impact of an SET}

\end{algorithm}

\section{\label{sec:Results}Results}

We perform our experiments on \textit{ISCAS'85 }(c432 etc)\textit{,
ISCAS'89 }(s344 etc) and \textit{ITC'99 }(b01 etc) benchmark circuits
in 180nm and 65nm technologies. Flip-flops (registers) are added to
the inputs and outputs of the combinational circuits of the \textit{ISCAS'85}
benchmarks. Clock frequency for each circuit is set to the maximum
operable frequency of the post-layout netlist, which is determined
by post-layout timing analysis. We calculate the probability of a
`gate-strike' and `register-strike' depending on whether the SET was
injected on a gate or a register. Further, we classify the bit-flips
into one of the SEU outcomes described in Figure \ref{fig:taxonomy}.
Our key focus is on quantifying the extent to which multiple flips
occur. We further evaluate the multiple bit flip probability contributed
independently by the `gate-strike' and `register strike' scenarios.

\subsection{\label{subsec:Probabilities-of-SET}Probabilities of SET outcomes}

\begin{figure}
\includegraphics[scale=0.45]{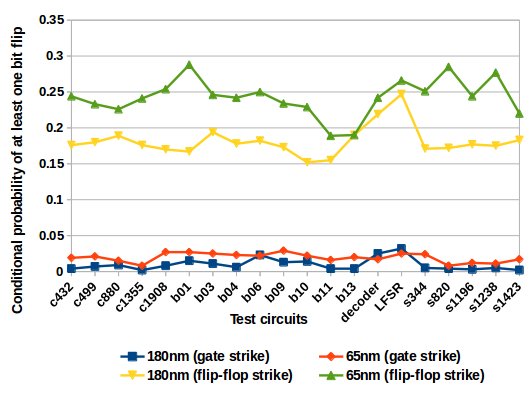}

\caption{\label{fig:Probability-of-one-}Probability of at least one bit-flip
for `gate-strike' and `register-strike' scenarios}

\end{figure}

In Table \ref{tab:SEU-flip-flop}, we show the observed probabilities
of all the SEU outcomes under the condition that the SET occurs at
a register (`register strike') as shown below. 
\begin{itemize}
\item $P_{NN}=P(NN|strike\:on\:register)$ 
\item $P_{NF}=P(NF|strike\:on\:register)$ 
\item $P_{FN}=P(FN|strike\:on\:register)$ 
\item $P_{FF}=P(FF|strike\:on\:register)$ 
\end{itemize}
In Table \ref{tab:SEU-gate}, we tabulate these SEU probabilities
under the condition that the SET occurs at a logic gate ( `gate-strike').
A `-' in the table means that the outcome did not occur. The key observations
are as follows:
\begin{itemize}
\item In both `register strike' and `gate strike' cases (Table \ref{tab:SEU-flip-flop}
and Table \ref{tab:SEU-gate}), $P_{NN}$, that is, probability of
no-flips is dominant, that is, the SET causes no flips. 
\item The probability $1-P_{NN}$ indicates the probability that a flip
or an error occurred. This is plotted in Figure \ref{fig:Probability-of-one-}.
The flip probability for register-strike cases is substantially greater
(nearly 10x greater) than that for the gate-strike cases. This indicates
that, when an SET occurs on a register, it is more likely to cause
an error, than one at a gate. Protecting the flip-flops or having
robust flip-flop designs can help reduce this flip probability.
\item From Figure \ref{fig:Probability-of-one-}, we note that the flip
probabilities are higher in the 65nm technology, which means that,
as technology scales, the likelihood of an error increases. 
\item In the case of strike on a register, from Table \ref{tab:SEU-flip-flop}
we observe that, $P_{FN}$ and $P_{FF}$ are dominant as compared
to $P_{NF}$. This implies that, when an SET occurs at a register,
it is more likely to cause a flip in the same clock cycle (clock cycle
`k'). 
\item On the other hand, when an SET occurs on a logic gate, from Table
\ref{tab:SEU-gate} we can see that the probabilities of $P_{FN}$
and $P_{FF}$ are extremely small as compared to $P_{NF}$. This indicates
that, when a strike occurs on a logic gate in clock cycle `k', the
register flip is most likely to occur due to the propagation of the
SET and hence, the flip is likely to occur in the subsequent clock
cycle `k+1'. 
\end{itemize}
\begin{table*}
\caption{\label{tab:SEU-flip-flop}SEU outcome probabilities when the SET occurs
at a register}

\begin{tabular}{>{\centering}p{1.4cm}|>{\centering}p{1.4cm}|>{\centering}p{1.4cm}|>{\centering}p{1.4cm}|>{\centering}p{1.4cm}|>{\centering}p{1.4cm}|>{\centering}p{1.4cm}|>{\centering}p{1.4cm}|>{\centering}p{1.4cm}}
\hline 
{\footnotesize{}Example}  & {\footnotesize{}$P_{NN}$ }{\footnotesize \par}

{\footnotesize{}180nm}  & {\footnotesize{}$P_{NN}$ }{\footnotesize \par}

{\footnotesize{}65nm}  & {\footnotesize{}$P_{NF}$ }{\footnotesize \par}

{\footnotesize{}180nm}  & {\footnotesize{}$P_{NF}$ }{\footnotesize \par}

{\footnotesize{}65nm}  & {\footnotesize{}$P_{FN}$ }{\footnotesize \par}

{\footnotesize{}180nm}  & {\footnotesize{}$P_{FN}$ }{\footnotesize \par}

{\footnotesize{}65nm}  & {\footnotesize{}$P_{FF}$ }{\footnotesize \par}

{\footnotesize{}180nm}  & {\footnotesize{}$P_{FF}$ }{\footnotesize \par}

{\footnotesize{}65nm}\tabularnewline
\hline 
\hline 
{\footnotesize{}c432}  & {\footnotesize{}0.824}  & {\footnotesize{}0.777}  & {\footnotesize{}0.003}  & {\footnotesize{}0.0275}  & {\footnotesize{}0.136}  & {\footnotesize{}0.166}  & {\footnotesize{}0.035}  & {\footnotesize{}0.028 }\tabularnewline
\hline 
{\footnotesize{}c499}  & {\footnotesize{}0.82}  & {\footnotesize{}0.767}  & {\footnotesize{}0.01}  & {\footnotesize{}0.029}  & {\footnotesize{}0.1}  & {\footnotesize{}0.113}  & {\footnotesize{}0.065}  & {\footnotesize{}0.089}\tabularnewline
\hline 
{\footnotesize{}c880}  & {\footnotesize{}0.811}  & {\footnotesize{}0.774}  & {\footnotesize{}0.005}  & {\footnotesize{}0.0225}  & {\footnotesize{}0.115}  & {\footnotesize{}0.135}  & {\footnotesize{}0.067}  & {\footnotesize{}0.067}\tabularnewline
\hline 
{\footnotesize{}c1355}  & {\footnotesize{}0.824}  & {\footnotesize{}0.759}  & {\footnotesize{}0.003}  & {\footnotesize{}0.027}  & {\footnotesize{}0.098}  & {\footnotesize{}0.11}  & {\footnotesize{}0.072}  & {\footnotesize{}0.102}\tabularnewline
\hline 
{\footnotesize{}c1908}  & {\footnotesize{}0.83}  & {\footnotesize{}0.746}  & {\footnotesize{}0.006}  & {\footnotesize{}0.037}  & {\footnotesize{}0.112}  & {\footnotesize{}0.137}  & {\footnotesize{}0.05}  & {\footnotesize{}0.078}\tabularnewline
\hline 
{\footnotesize{}b01}  & {\footnotesize{}0.833}  & {\footnotesize{}0.712}  & {\footnotesize{}0.013}  & {\footnotesize{}0.038}  & {\footnotesize{}0.088}  & {\footnotesize{}0.116}  & {\footnotesize{}0.064}  & {\footnotesize{}0.133}\tabularnewline
\hline 
{\footnotesize{}b03}  & {\footnotesize{}0.806}  & {\footnotesize{}0.7583}  & {\footnotesize{}0.006}  & {\footnotesize{}0.034}  & {\footnotesize{}0.046}  & {\footnotesize{}0.065}  & {\footnotesize{}0.14}  & {\footnotesize{}0.141}\tabularnewline
\hline 
{\footnotesize{}b04}  & {\footnotesize{}0.822}  & {\footnotesize{}0.808}  & {\footnotesize{}0.002}  & {\footnotesize{}0.012}  & {\footnotesize{}0.05}  & {\footnotesize{}0.057}  & {\footnotesize{}0.124}  & {\footnotesize{}0.122}\tabularnewline
\hline 
{\footnotesize{}b06}  & {\footnotesize{}0.818}  & {\footnotesize{}0.755}  & {\footnotesize{}0.0128}  & {\footnotesize{}0.024}  & {\footnotesize{}0.117}  & {\footnotesize{}0.156}  & {\footnotesize{}0.051}  & {\footnotesize{}0.063}\tabularnewline
\hline 
{\footnotesize{}b09}  & {\footnotesize{}0.827}  & {\footnotesize{}0.766}  & {\footnotesize{}0.006}  & {\footnotesize{}0.025}  & {\footnotesize{}0.024}  & {\footnotesize{}0.023}  & {\footnotesize{}0.141}  & {\footnotesize{}0.183}\tabularnewline
\hline 
{\footnotesize{}b10}  & {\footnotesize{}0.848}  & {\footnotesize{}0.771}  & {\footnotesize{}0.003}  & {\footnotesize{}0.029}  & {\footnotesize{}0.011}  & {\footnotesize{}0.022}  & {\footnotesize{}0.136}  & {\footnotesize{}0.177}\tabularnewline
\hline 
{\footnotesize{}b11}  & {\footnotesize{}0.845}  & {\footnotesize{}0.811}  & {\footnotesize{}0.006}  & {\footnotesize{}0.022}  & {\footnotesize{}0.025}  & {\footnotesize{}0.046}  & {\footnotesize{}0.121}  & {\footnotesize{}0.119}\tabularnewline
\hline 
{\footnotesize{}b13}  & {\footnotesize{}0.81}  & {\footnotesize{}0.81}  & {\footnotesize{}0.008}  & {\footnotesize{}0.026}  & {\footnotesize{}0.012}  & {\footnotesize{}0.013}  & {\footnotesize{}0.168}  & {\footnotesize{}0.146}\tabularnewline
\hline 
\textcolor{black}{\footnotesize{}s344} & \textcolor{black}{\footnotesize{}0.829} & \textcolor{black}{\footnotesize{}0.749} & \textcolor{black}{\footnotesize{}0.011} & \textcolor{black}{\footnotesize{}0.046} & \textcolor{black}{\footnotesize{}0.136} & \textcolor{black}{\footnotesize{}0.17} & \textcolor{black}{\footnotesize{}0.024} & \textcolor{black}{\footnotesize{}0.033}\tabularnewline
\hline 
\textcolor{black}{\footnotesize{}s820} & \textcolor{black}{\footnotesize{}0.828} & \textcolor{black}{\footnotesize{}0.715} & \textcolor{black}{\footnotesize{}0.024} & \textcolor{black}{\footnotesize{}0.022} & \textcolor{black}{\footnotesize{}0.115} & \textcolor{black}{\footnotesize{}0.159} & \textcolor{black}{\footnotesize{}0.031} & \textcolor{black}{\footnotesize{}0.1}\tabularnewline
\hline 
\textcolor{black}{\footnotesize{}s1196} & \textcolor{black}{\footnotesize{}0.823} & \textcolor{black}{\footnotesize{}0.756} & \textcolor{black}{\footnotesize{}0.014} & \textcolor{black}{\footnotesize{}0.025} & \textcolor{black}{\footnotesize{}0.155} & \textcolor{black}{\footnotesize{}0.2} & \textcolor{black}{\footnotesize{}0.008} & \textcolor{black}{\footnotesize{}0.018}\tabularnewline
\hline 
\textcolor{black}{\footnotesize{}s1238} & \textcolor{black}{\footnotesize{}0.825} & \textcolor{black}{\footnotesize{}0.723} & \textcolor{black}{\footnotesize{}0.007} & \textcolor{black}{\footnotesize{}0.039} & \textcolor{black}{\footnotesize{}0.157} & \textcolor{black}{\footnotesize{}0.219} & \textcolor{black}{\footnotesize{}0.01} & \textcolor{black}{\footnotesize{}0.018}\tabularnewline
\hline 
\textcolor{black}{\footnotesize{}s1423} & \textcolor{black}{\footnotesize{}0.817} & \textcolor{black}{\footnotesize{}0.78} & \textcolor{black}{\footnotesize{}0.004} & \textcolor{black}{\footnotesize{}0.026} & \textcolor{black}{\footnotesize{}0.112} & \textcolor{black}{\footnotesize{}0.103} & \textcolor{black}{\footnotesize{}0.066} & \textcolor{black}{\footnotesize{}0.089}\tabularnewline
\hline 
{\footnotesize{}3:8 decoder}  & {\footnotesize{}0.781}  & {\footnotesize{}0.732}  & {\footnotesize{}0.021}  & {\footnotesize{}0.041}  & {\footnotesize{}0.156}  & {\footnotesize{}0.154}  & {\footnotesize{}0.04}  & {\footnotesize{}0.072}\tabularnewline
\hline 
{\footnotesize{}8-bit LFSR}  & {\footnotesize{}0.753}  & {\footnotesize{}0.734}  & {\footnotesize{}0.038}  & {\footnotesize{}0.034}  & {\footnotesize{}0.012}  & {\footnotesize{}0.01}  & {\footnotesize{}0.195}  & {\footnotesize{}0.22 }\tabularnewline
\hline 
{\footnotesize{}worst case standard error}  & {\footnotesize{}$\pm0.01$}  & {\footnotesize{}$\pm0.02$}  & {\footnotesize{}$\pm0.009$}  & {\footnotesize{}$\pm0.008$}  & {\footnotesize{}$\pm0.016$}  & {\footnotesize{}$\pm0.018$}  & {\footnotesize{}$\pm0.015$}  & {\footnotesize{}$\pm0.017$}\tabularnewline
\hline 
\end{tabular}
\end{table*}
\begin{table*}
\caption{\label{tab:SEU-gate}SEU outcome probabilities when the SET occurs
at a logic gate (`-' indicates no events were observed for that particular
case)}

\begin{tabular}{>{\centering}p{1.4cm}|>{\centering}p{1.4cm}|>{\centering}p{1.4cm}|>{\centering}p{1.4cm}|>{\centering}p{1.4cm}|>{\centering}p{1.4cm}|>{\centering}p{1.4cm}|>{\centering}p{1.4cm}|>{\centering}p{1.4cm}}
\hline 
{\footnotesize{}Example}  & {\footnotesize{}$P_{NN}$ }{\footnotesize \par}

{\footnotesize{}180nm}  & {\footnotesize{}$P_{NN}$ }{\footnotesize \par}

{\footnotesize{}65nm}  & {\footnotesize{}$P_{NF}$ }{\footnotesize \par}

{\footnotesize{}180nm}  & {\footnotesize{}$P_{NF}$ }{\footnotesize \par}

{\footnotesize{}65nm}  & {\footnotesize{}$P_{FN}$ }{\footnotesize \par}

{\footnotesize{}180nm}  & {\footnotesize{}$P_{FN}$ }{\footnotesize \par}

{\footnotesize{}65nm}  & {\footnotesize{}$P_{FF}$ }{\footnotesize \par}

{\footnotesize{}180nm}  & {\footnotesize{}$P_{FF}$ }{\footnotesize \par}

{\footnotesize{}65nm}\tabularnewline
\hline 
\hline 
{\footnotesize{}c432}  & {\footnotesize{}0.996}  & {\footnotesize{}0.98}  & {\footnotesize{}0.003}  & {\footnotesize{}0.019}  & -  & {\footnotesize{}-}  & {\footnotesize{}-}  & {\footnotesize{}- }\tabularnewline
\hline 
{\footnotesize{}c499}  & {\footnotesize{}0.993}  & {\footnotesize{}0.979}  & {\footnotesize{}0.004}  & {\footnotesize{}0.019}  & {\footnotesize{}0.001}  & {\footnotesize{}0.0008}  & {\footnotesize{}-}  & {\footnotesize{}-}\tabularnewline
\hline 
{\footnotesize{}c880}  & {\footnotesize{}0.991}  & {\footnotesize{}0.985}  & {\footnotesize{}0.006}  & {\footnotesize{}0.013}  & {\footnotesize{}0.002}  & -  & -  & {\footnotesize{}0.0016}\tabularnewline
\hline 
{\footnotesize{}c1355}  & {\footnotesize{}0.998}  & {\footnotesize{}0.992}  & {\footnotesize{}0.0016}  & {\footnotesize{}0.007}  & -  & {\footnotesize{}-}  & -  & {\footnotesize{}-}\tabularnewline
\hline 
{\footnotesize{}c1908}  & {\footnotesize{}0.992}  & {\footnotesize{}0.973}  & {\footnotesize{}0.007}  & {\footnotesize{}0.025}  & -  & {\footnotesize{}-}  & -  & {\footnotesize{}0.0005}\tabularnewline
\hline 
{\footnotesize{}b01}  & {\footnotesize{}0.985}  & {\footnotesize{}0.973}  & {\footnotesize{}0.014}  & {\footnotesize{}0.026}  & -  & {\footnotesize{}-}  & -  & {\footnotesize{}- }\tabularnewline
\hline 
{\footnotesize{}b03}  & {\footnotesize{}0.989}  & {\footnotesize{}0.965}  & {\footnotesize{}0.006}  & {\footnotesize{}0.034}  & -  & {\footnotesize{}-}  & {\footnotesize{}0.003}  & {\footnotesize{}-}\tabularnewline
\hline 
{\footnotesize{}b04}  & {\footnotesize{}0.994}  & {\footnotesize{}0.977}  & {\footnotesize{}0.002}  & {\footnotesize{}0.021}  & {\footnotesize{}0.0005}  & {\footnotesize{}0.001}  & {\footnotesize{}0.002}  & {\footnotesize{}-}\tabularnewline
\hline 
{\footnotesize{}b06}  & {\footnotesize{}0.977}  & {\footnotesize{}0.981}  & {\footnotesize{}0.021}  & {\footnotesize{}0.017}  & {\footnotesize{}0.0004}  & {\footnotesize{}0.0005}  & -  & -\tabularnewline
\hline 
{\footnotesize{}b09}  & {\footnotesize{}0.987}  & {\footnotesize{}0.971}  & {\footnotesize{}0.008}  & {\footnotesize{}0.028}  & {\footnotesize{}0.0005}  & {\footnotesize{}-}  & {\footnotesize{}0.003}  & {\footnotesize{}- }\tabularnewline
\hline 
{\footnotesize{}b10}  & {\footnotesize{}0.986}  & {\footnotesize{}0.978}  & {\footnotesize{}0.011}  & {\footnotesize{}0.021}  & {\footnotesize{}-}  & {\footnotesize{}-}  & {\footnotesize{}0.002}  & {\footnotesize{}-}\tabularnewline
\hline 
{\footnotesize{}b11}  & {\footnotesize{}0.996}  & {\footnotesize{}0.984}  & {\footnotesize{}0.001}  & {\footnotesize{}0.015}  & {\footnotesize{}-}  & {\footnotesize{}-}  & {\footnotesize{}0.002}  & {\footnotesize{}-}\tabularnewline
\hline 
{\footnotesize{}b13}  & {\footnotesize{}0.996}  & {\footnotesize{}0.98}  & {\footnotesize{}0.011}  & {\footnotesize{}0.019}  & -  & -  & {\footnotesize{}0.0026}  & -\tabularnewline
\hline 
\textcolor{black}{\footnotesize{}s344} & \textcolor{black}{\footnotesize{}0.995} & \textcolor{black}{\footnotesize{}0.976} & \textcolor{black}{\footnotesize{}0.004} & \textcolor{black}{\footnotesize{}0.022} & \textcolor{black}{\footnotesize{}-} & \textcolor{black}{\footnotesize{}0.001} & - & -\tabularnewline
\hline 
\textcolor{black}{\footnotesize{}s820} & \textcolor{black}{\footnotesize{}0.996} & \textcolor{black}{\footnotesize{}0.992} & \textcolor{black}{\footnotesize{}0.0022} & \textcolor{black}{\footnotesize{}0.007} & \textcolor{black}{\footnotesize{}0.0015} & \textcolor{black}{\footnotesize{}-} & {\footnotesize{}- } & {\footnotesize{}- }\tabularnewline
\hline 
\textcolor{black}{\footnotesize{}s1196} & \textcolor{black}{\footnotesize{}0.997} & \textcolor{black}{\footnotesize{}0.988} & \textcolor{black}{\footnotesize{}0.003} & \textcolor{black}{\footnotesize{}0.01} & - & \textcolor{black}{\footnotesize{}0.001} & {\footnotesize{}-} & {\footnotesize{}-}\tabularnewline
\hline 
\textcolor{black}{\footnotesize{}s1238} & \textcolor{black}{\footnotesize{}0.995} & \textcolor{black}{\footnotesize{}0.989} & \textcolor{black}{\footnotesize{}0.005} & \textcolor{black}{\footnotesize{}0.01} & {\footnotesize{}- } & \textcolor{black}{\footnotesize{}-} & {\footnotesize{}-} & {\footnotesize{}-}\tabularnewline
\hline 
\textcolor{black}{\footnotesize{}s1423} & \textcolor{black}{\footnotesize{}0.998} & \textcolor{black}{\footnotesize{}0.983} & \textcolor{black}{\footnotesize{}0.002} & \textcolor{black}{\footnotesize{}0.015} & {\footnotesize{}-} & \textcolor{black}{\footnotesize{}0.001} & - & -\tabularnewline
\hline 
{\footnotesize{}3:8 decoder}  & {\footnotesize{}0.975}  & {\footnotesize{}0.989}  & {\footnotesize{}0.0185}  & {\footnotesize{}0.01}  & {\footnotesize{}0.006}  & {\footnotesize{}-}  & -  & {\footnotesize{}-}\tabularnewline
\hline 
{\footnotesize{}8-bit LFSR}  & {\footnotesize{}0.968}  & {\footnotesize{}0.975}  & {\footnotesize{}0.029}  & {\footnotesize{}0.023}  & -  & -  & {\footnotesize{}0.001}  & {\footnotesize{}0.0005 }\tabularnewline
\hline 
{\footnotesize{}worst case standard error}  & {\footnotesize{}$\pm0.009$}  & {\footnotesize{}$\pm0.009$}  & {\footnotesize{}$\pm0.009$}  & {\footnotesize{}$\pm0.008$}  & {\footnotesize{}$\pm0.004$}  & {\footnotesize{}$\pm0.001$}  & {\footnotesize{}$\pm0.003$}  & {\footnotesize{}$\pm0.002$}\tabularnewline
\hline 
\end{tabular}
\end{table*}

\subsection{\label{subsec:Impact-of-a}Probability of multiple register flips
given an erroneous outcome}

\textcolor{black}{Given that an error or a flip occurred due to an
SET, that is, }one of these cases occurred- $NF^{*}$, $F^{*}N$ or
$FF^{*}$\textcolor{black}{{} (}where $F^{*}$ means one or more flips)\textcolor{black}{,
what is the likelihood that the first error event caused by the SET
affects multiple registers? There are two possibilities: }
\begin{itemize}
\item \textcolor{black}{The first error event occurs at cycle $k+1$ and
consists of multiple registers being in error, that is, }$P_{m}=P(NF_{m}|atleast\,one\,flip)$.\textcolor{black}{{} }
\item \textcolor{black}{The first error event occurs at cycle $k$ and consists
of multiple registers being in error, that is,} $P(F_{m}N|atleast\,one\,flip)$.\textcolor{black}{{}
There is another scenario where the multiple flips leads to flip in
the next cycle:} $P(F_{m}F|atleast\,one\,flip)$.\textcolor{black}{{} }
\end{itemize}
We do not consider the probability $P(FF_{m}|atleast\,one\,flip)$,
because the multiple flips in cycle \textcolor{black}{$k+1$} are
caused due to the propagation of the first flip, which will be anyway
captured by modeling the first flip using the traditional single bit-flip
model. So, the multiple flips in this case are not a direct result
of the SET.

Amongst the two cases mentioned above, we did not come across the
scenario of $P(F_{m}N|atleast\,one\,flip)$ or $P(F_{m}F|atleast\,one\,flip)$
in our experiments, that is, multiple flips in clock cycle `k'. Hence,
we report the first probability: $P_{m}=P(NF_{m}|atleast\,one\,flip)$.
This is calculated as follows for both the gate-strike and register-strike
(denoted by reg-strike in the equation) cases put together:

\begin{equation}
P_{m}=\frac{No.\:of\:(NF_{m})_{Gate-strike}+No.\:of\:(NF_{m})_{reg-strike}}{(No.\:of\:cases\:with\:atleast\:one\:flip)_{gate+reg-strike}}
\end{equation}

In other words, we calculate

\begin{equation}
P_{m}=\frac{No.\:of\:(NF_{m})_{Gate-strike}+No.\:of\:(NF_{m})_{reg-strike}}{(No.\:of\:NF,\:FN\:or\:FF)_{gate+reg-strike}}
\end{equation}

For instance, if we ran 5000 simulations, out of which there were
1000 cases in which at least one error occurred (this can include
$NF$, $FN$ or $FF$ across gate and register strikes), out of which,
say 100 cases had multiple errors ($NF_{m}$), we calculate $P_{m}$
as 100/1000. This probability is shown in Table \textcolor{black}{\ref{tab:Multiple-flips-gate_reg}}
for the benchmark circuits, across both register and gate strikes.

\begin{table}[h]
\caption{\label{tab:Multiple-flips-1}Probability of multiple bit flips given
that at least one error occurred across all strikes (`-' indicates
no multiple error events were observed)}

$\;\;\;\;\;\;\;\;\;\;\;\;$%
\begin{tabular}{>{\centering}p{1.3cm}|>{\centering}p{1.2cm}|>{\centering}p{1.2cm}}
\hline 
{\footnotesize{}Example}  & $P_{m}$ {\footnotesize{}-}{\footnotesize \par}

{\footnotesize{}`gate and register strike'}{\footnotesize \par}

{\footnotesize{}180nm}  & $P_{m}$-

{\footnotesize{}`gate and register strike'}{\footnotesize \par}

{\footnotesize{}65nm}\tabularnewline
\hline 
\hline 
{\footnotesize{}c432}  & {\footnotesize{}0.006} & {\footnotesize{}0.035}\tabularnewline
\hline 
{\footnotesize{}c499}  & {\footnotesize{}0.003} & {\footnotesize{}0.016}\tabularnewline
\hline 
{\footnotesize{}c880}  & -  & {\footnotesize{}0.001}\tabularnewline
\hline 
{\footnotesize{}c1355}  & {\footnotesize{}0.003} & {\footnotesize{}0.007}\tabularnewline
\hline 
{\footnotesize{}c1908}  & {\footnotesize{}0.003}  & {\footnotesize{}0.037}\tabularnewline
\hline 
{\footnotesize{}b01}  & -  & {\footnotesize{}0.02}\tabularnewline
\hline 
{\footnotesize{}b03}  & {\footnotesize{}0.005}  & {\footnotesize{}0.007}\tabularnewline
\hline 
{\footnotesize{}b04}  & -  & {\footnotesize{}0.03}\tabularnewline
\hline 
{\footnotesize{}b06}  & \textbf{\footnotesize{}0.042} & {\footnotesize{}0.05}\tabularnewline
\hline 
{\footnotesize{}b09}  & {\footnotesize{}0.005}  & {\footnotesize{}0.044}\tabularnewline
\hline 
{\footnotesize{}b10}  & {\footnotesize{}0.007}  & {\footnotesize{}0.06}\tabularnewline
\hline 
{\footnotesize{}b11}  & -  & {\footnotesize{}0.036}\tabularnewline
\hline 
{\footnotesize{}b13}  & -  & {\footnotesize{}0.038}\tabularnewline
\hline 
\textcolor{black}{\footnotesize{}s344} & -  & \textcolor{black}{\footnotesize{}0.02}\tabularnewline
\hline 
\textcolor{black}{\footnotesize{}s820} & -  & \textbf{\textcolor{black}{\footnotesize{}0.081}}\tabularnewline
\hline 
\textcolor{black}{\footnotesize{}s1196} & -  & \textcolor{black}{\footnotesize{}0.014}\tabularnewline
\hline 
\textcolor{black}{\footnotesize{}s1238} & -  & \textcolor{black}{\footnotesize{}0.022}\tabularnewline
\hline 
\textcolor{black}{\footnotesize{}s1423} & - & \textcolor{black}{\footnotesize{}0.028}\tabularnewline
\hline 
{\footnotesize{}3:8 decoder}  & {\footnotesize{}0.006}  & {\footnotesize{}0.014}\tabularnewline
\hline 
{\footnotesize{}8-bit LFSR}  & {\footnotesize{}0.007}  & {\footnotesize{}0.018}\tabularnewline
\hline 
\end{tabular}
\end{table}

From Table \textcolor{black}{\ref{tab:Multiple-flips-gate_reg}},
we see that the maximum probability of multiple flips $P_{m}$ that
was observed in the circuits we simulated, across both register and
gate strikes was $8\%$. This probability is significantly higher
in most of the 65nm implementations as compared to the 180nm implementations.
This probability is likely to depend on several factors such as the
number of flip-flops, presence of balanced paths, logic depth, logic
gates, input combinations and so on. Although 8\% seems low, it can
still have a significant impact on error detection and correction
schemes. This means that, a single error correction scheme can go
wrong 8\% of the times and the errors can go down as silent undetected
errors.

\subsection{The contribution of gate strikes and register strikes to multiple
register flips}

In this section, we analyze our results further to understand the
role of gate-strikes and register-strikes independently on the multiple
flip probability. In other words, we address the following question.
Given that an error occurred due to a strike on a logic gate, what
is the probability that it was a multiple flip? This is denoted by
$P_{GM}$. Similarly, we calculate the multiple flip probability $P_{RM}$
given that an error was caused due to a register strike. We calculate
$P_{GM}$ and $P_{RM}$ respectively as follows:

\begin{equation}
P_{GM}=\frac{Number\:of\:(NF_{m})_{gate-strike}}{(Number\:of\:NF,\:FN\:or\:FF)_{gate-strike}}
\end{equation}

\begin{equation}
P_{RM}=\frac{Number\:of\:(NF_{m})_{register-strike}}{(Number\:of\:NF,\:FN\:or\:FF)_{register-strike}}
\end{equation}

In these equations, we should note that, the denominator in both the
fractions is the total number of cases which had erroneous outcomes
(flips). For the gate strike scenario, when an SET occurs on a logic
gate in clock cycle `k', the register flip is most likely to occur
due to the propagation of the SET and hence, the flip is likely to
occur in the subsequent clock cycle `k+1'. Thus, a major contributor
of the erroneous outcomes is the \textquoteleft NF\textquoteright{}
scenario. On the other and, for the register strike case, an SET is
more likely to cause an error in the same clock cycle and hence errors
are mainly contributed by the \textquoteleft FN\textquoteright{} and
\textquoteleft FF\textquoteright{} scenarios. These are already observed
in Tables \ref{tab:SEU-flip-flop} and \ref{tab:SEU-gate}. So, $P_{FN}$
and $P_{FF}$ are dominant for register strike cases and $P_{NF}$
is dominant for gate strike scenarios. These observations are seen
to have a significant impact on the probabilities $P_{GM}$ and $P_{RM}$.
We present these probabilities in Table \ref{tab:Multiple-flips-gate_reg}.
We find that $P_{GM}$ is significantly greater than $P_{RM}$ and
its value is found to be a maximum of 50\%. This is mainly because
of two reasons:
\begin{itemize}
\item When an SET occurs on a register, 

\begin{itemize}
\item The probability of a flip or an error occuring is high (high probabilities
of an $FF$ or an $FN$) as compared to the SET occurring on a gate
(Figure \ref{fig:Probability-of-one-}). The flip probability for
register-strike cases is found to be nearly 10x greater than that
for the gate-strike cases. This increases the denominator in equation
4 and results in a small value for $P_{RM}$.
\item The probability of an $NF$ scenario is small, as compared to the
$FN$ and $FF$ scenarios for a register strike as already observed
in Table \ref{tab:SEU-flip-flop}. Hence $NF_{m}$ scenarios are also
less leading to a small value in the numerator in equation 4. This
again leads to low value of $P_{RM}$.
\end{itemize}
\item When an SET occurs on a logic gate, 

\begin{itemize}
\item The probability of a flip is low (low probabilities of an $FF$ or
an $FN$) as compared to the SET occurring on a register as already
observed in Figure \ref{fig:Probability-of-one-}. This decreases
the denominator in equation 3.
\item The major contributor of flips for gate strikes is $NF$ (Table \ref{tab:SEU-gate}).
Hence $NF_{m}$ scenarios are also high, as compared to the register
strike cases, leading to a large value in the numerator in equation
3. These lead to a high value of $P_{GM}$.
\end{itemize}
\end{itemize}
\begin{table}
\caption{\label{tab:Multiple-flips-gate_reg}Probability of multiple bit flips
given that at least one error occurred. Gate strikes and register
strikes are presented separately. (`-' indicates no multiple error
events were observed)}

\begin{tabular}{>{\centering}p{1.5cm}|>{\centering}p{1.3cm}|>{\centering}p{1.3cm}|>{\centering}p{1.3cm}|>{\centering}p{1.3cm}}
\hline 
{\footnotesize{}Example}  & $P_{GM}$ {\footnotesize{}-}{\footnotesize \par}

{\footnotesize{}`gate strike'}{\footnotesize \par}

{\footnotesize{}180nm}  & $P_{GM}$ {\footnotesize{}-}{\footnotesize \par}

{\footnotesize{}`gate strike'}{\footnotesize \par}

{\footnotesize{}65nm}  & $P_{RM}$ {\footnotesize{}-}{\footnotesize \par}

{\footnotesize{}`register strike'}{\footnotesize \par}

{\footnotesize{}180nm}  & $P_{RM}$ {\footnotesize{}-}{\footnotesize \par}

{\footnotesize{}`register strike'}{\footnotesize \par}

{\footnotesize{}65nm} \tabularnewline
\hline 
\hline 
{\footnotesize{}c432}  & {\footnotesize{}0.4} & \textbf{\footnotesize{}0.5} & -  & {\footnotesize{}0.002}\tabularnewline
\hline 
{\footnotesize{}c499}  & {\footnotesize{}0.11} & {\footnotesize{}0.2} & -  & -\tabularnewline
\hline 
{\footnotesize{}c880}  & -  & - & -  & -\tabularnewline
\hline 
{\footnotesize{}c1355}  & \textbf{\footnotesize{}0.33} & {\footnotesize{}0.2} & -  & -\tabularnewline
\hline 
{\footnotesize{}c1908}  & {\footnotesize{}0.067} & {\footnotesize{}0.25} & -  & {\footnotesize{}0.008}\tabularnewline
\hline 
{\footnotesize{}b01}  & -  & {\footnotesize{}0.035} & -  & \textbf{\footnotesize{}0.016}\tabularnewline
\hline 
{\footnotesize{}b03}  & {\footnotesize{}0.12} & {\footnotesize{}0.042} & -  & {\footnotesize{}0.004}\tabularnewline
\hline 
{\footnotesize{}b04}  & -  & {\footnotesize{}0.172} & -  & -\tabularnewline
\hline 
{\footnotesize{}b06}  & \textbf{\footnotesize{}0.288} & {\footnotesize{}0.35} & -  & {\footnotesize{}0.003}\tabularnewline
\hline 
{\footnotesize{}b09}  & {\footnotesize{}0.08} & {\footnotesize{}0.36} & -  & {\footnotesize{}0.003}\tabularnewline
\hline 
{\footnotesize{}b10}  & {\footnotesize{}0.058} & \textbf{\footnotesize{}0.4} & -  & -\tabularnewline
\hline 
{\footnotesize{}b11}  & -  & {\footnotesize{}0.12} & -  & {\footnotesize{}0.0012}\tabularnewline
\hline 
{\footnotesize{}b13}  & -  & {\footnotesize{}0.34} & -  & -\tabularnewline
\hline 
\textcolor{black}{\footnotesize{}s344} & -  & \textcolor{black}{\footnotesize{}0.07} & -  & \textcolor{black}{\footnotesize{}0.012}\tabularnewline
\hline 
\textcolor{black}{\footnotesize{}s820} & -  & \textcolor{black}{\footnotesize{}0.12} & -  & -\tabularnewline
\hline 
\textcolor{black}{\footnotesize{}s1196} & -  & \textcolor{black}{\footnotesize{}0.045} & -  & -\tabularnewline
\hline 
\textcolor{black}{\footnotesize{}s1238} & -  & \textcolor{black}{\footnotesize{}0.08} & -  & -\tabularnewline
\hline 
\textcolor{black}{\footnotesize{}s1423} & - & \textcolor{black}{\footnotesize{}0.27} & - & \textcolor{black}{\footnotesize{}0.003}\tabularnewline
\hline 
{\footnotesize{}3:8 decoder}  & {\footnotesize{}0.107} & {\footnotesize{}0.27} & \textbf{\footnotesize{}0.001} & -\tabularnewline
\hline 
{\footnotesize{}8-bit LFSR}  & {\footnotesize{}0.073} & {\footnotesize{}0.17} & {\footnotesize{}-}  & -\tabularnewline
\hline 
\end{tabular}
\end{table}

\begin{figure}
\includegraphics[scale=0.4]{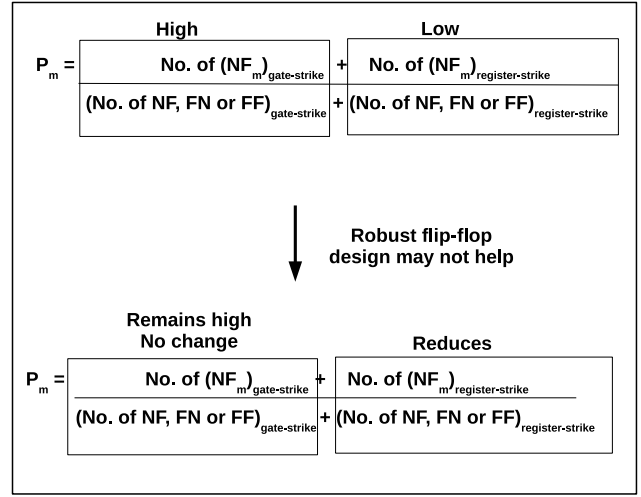}

\caption{\label{fig:Why-a-robust}Why a robust flip-flop design will not help
reduce multiple bit-flip probability?}
\end{figure}

To summarize, given that there is an error, the probability of multiple
errors in the case of register-strike cases is less: about 2\%, where
as they are significant in the gate-strike cases and can be up to
50\%. Thus, in a circuit design with robust flip-flops, we can eliminate
the flips due to `register strike' cases, but out of the remaining
flips which are going to be due to `gate strikes', multiple flips
will be extremely likely. This is depicted in Figure \ref{fig:Why-a-robust}.
Thus, gate strikes are more likely to be problematic. Hardening the
flip-flops is not going to help in this scenario. Alternate methods
such as delay capture flip-flop designs or uneven path delays may
need to be used to prevent multiple flips due to gate strikes.

\section{\label{sec:Conclusions}Conclusions}

We have performed a detailed characterization of the impact of an
SET on a logic circuit to quantify the extent to which the single
bit-flip model is realistic. Our goal was to evaluate the multiple
bit-flip error probability. Our observation is that, in up to 8\%
of the outcomes when an error occurred, an SET created multiple register
errors, that is, the single bit-flip model is optimistic in up to
8\% of the cases. Although this probability may seem low, it can have
a significant impact on error-detection or correction schemes, that
is, a single-bit error correction scheme can go wrong in up to 8\%
of the times and these errors will remain as silent undetected errors.
Further, the probability of multiple errors increases as technology
is scaled. A key observation is that, amongst the erroneous outcomes,
the probability of multiple bit flips for gate strike cases was substantial
and went up to 50\%, where as those for register strike cases was
just about 2\%. This implies that, in principle, we can eliminate
the flips due to register strikes using robust flip-flop designs.
But in such designs, out of the remaining flips which will be due
to gate strikes, a large fraction will be multiple flips. Thus, there
is a need to focus on circuit techniques to eliminate these type of
errors.

\bibliographystyle{ieeetr}
\bibliography{paper_cktsim_gates}

\end{document}